\documentstyle[12pt,moriond0,epsfig]{article}

\def\Journal#1#2#3#4{{#1} {\bf #2}, #3 (#4)}

\def\PLB{{\em Phys. Lett.}  B}

\def\PRD{{\em Phys. Rev.} D}

\newcommand{\comm}[1]{}

\newcommand{\trh}{T_{\rm rh}}
\newcommand{\teff}{T_{\rm eff}}

\newcommand{\ncs}{N_{_{\rm CS}}}

\newcommand{\phii}{\langle \phi^*\phi \rangle}

\newcommand{\ssqangle}{\langle\sigma^2\rangle}
\newcommand{\mue}{\mu_{\rm eff}}
\newcommand{\ncsa}{\langle N_{\rm CS}\rangle}

\newcommand{\ncst}{\langle N_{\rm CS}(t)\rangle}

\newcommand{\dncs}{\delta N_{\rm CS}}

\newcommand{\esph}{E_{\rm sph}}
\newcommand{\gsph}{\Gamma_{\rm sph}}
\newcommand{\be}{\begin{equation}}
\newcommand{\ee}{\end{equation}}
\newcommand{\sign}{\mathop{\rm sign}\nolimits}
\begin{document}

\thispagestyle{empty}

\def\thefootnote{\fnsymbol{footnote}}

\vspace*{4cm} \title{ELECTROWEAK BARYOGENESIS AT AND AFTER PREHEATING:
WHAT'S THE DIFFERENCE?
\footnote{To appear in the Proceedings of XXXVth Rencontres de
Moriond: Electroweak Interactions and Unified Theories, Les Arcs,
France, 11-18 March 2000.} }

\author{Dmitri Grigoriev
\footnote{e-mail: dgr@inr.ac.ru}}

\address{Institute for Nuclear Research of Russian Academy of
Sciences,\\
60th October Anniversary Prospect, 7a, Moscow 117312, Russia}

\maketitle

\abstracts{Two recent scenarios of preheating-related
baryogenesis are compared within the framework of Abelian Higgs model
in (1+1) dimensions. It is shown that they shift baryon number in
opposite directions. Once both scenarios can realize simultaneously as
overlapped stages of the same physical process, even the sign of net
generated asymmetry becomes dependent on initial parameters.}

\newpage

As it was shown recently, parametric resonance during preheating opens
new possibilities for electroweak baryogenesis \cite{krs,rs}  which can
take place both during~\cite{bgks} and after~\cite{bg} the
resonance. While baryogenesis mechanisms described in ~\cite{bgks} and
~\cite{bg} can appear in the same theory and even overlap in the
course of the same parametric resonance, they are based on different
dynamic effects. In the present paper we compare baryoproducing
efficiency of both mechanisms and discuss underlying dynamical
processes. The key difference between these mechanisms is that they
are related to different time scales existing in the model.
Baryogenesis at preheating exists on rather short parametric resonance
time scale, while baryogenesis after preheating is related to inflaton
thermalization time scale, that can become very large, at least in
classical dynamics~\cite{gq98}. The third time scale is the Higgs field
thermalization time which generally lies between the two previous time
scales. This intermediate time scale defines the freeze-out time of
post-resonant sphaleron transitions and is important because sphaleron
transitions may wash out the generated asymmetry if they'll keep going
after the end of baryoproduction. While the wash-out processes as well
as other fermion backreaction effects are beyond the scope of our
numerical simulations, the very existence of intensive sphaleron
transitions on the time scale considerably larger than baryoproduction
time is a signal for possible problems.

The resonance-localized baryogenesis~\cite{bgks} can rapidly produce
considerable shift in topological number. Unlike baryogenesis at
first-order phase transition, it occurs during extended period of time
on the resonance time scale, thus making the wash-out problem a bit
easier to solve. However, in this case wash-out suppression requires
rather short Higgs thermalization time which should be close to
resonance time scale. An attempt to achieve this in (1+1)-dimensional
theory being simulated in~\cite{bgks} leads to unacceptably high
reheating temperatures $\trh \sim 0.3\,v$ where sphaleron
transitions do not freeze out at all, see Fig.~8 of \cite{bgks}. Even
though this conjecture isn't directly applicable to realistic
(3+1)-dimensional case, it shows that more efficient ways to prevent
wash-out are highly desirable.

A promising approach was suggested in paper~\cite{bg} where it was
shown that for certain parameter values baryogenesis keeps going much
longer than the resonance itself. In fact, CP violation that drives
baryoproduction is present here for very long time, and
baryoproduction keeps going as long as the sphaleron transitions
occur. This scenario is based on difference between inflaton and Higgs
thermalization times. In our simulations Higgs field thermalizes much
faster than inflaton field because it couples to gauge field and has
nonlinear self-potential, which results in intensive rescattering of
Higgs spectral modes. As we will show below, the presence of
non-thermalized oscillating inflaton field can generate effective CP
violation, which exists on the inflaton thermalization time scale, the
longest one in the model. Of course, this CP violation can change the
baryon number only via sphaleron transitions, so the baryogenesis
actually occurs on Higgs thermalization time scale, freezing out
together with the sphaleron transitions. The major advantage of
approach~\cite{bg} is that the baryoproduction-driving effective
chemical potential exists here long after the sphaleron transitions
vanish. In other words, it gives us a toy model of electroweak
baryogenesis which is completely free from wash-out.

An interesting peculiarity of baryogenesis at and after preheating is
that both these scenarios can realize simultaneously, see
Fig.~\ref{smallres}.  Once for the same initial parameters they
generate baryonic asymmetry of opposite signs, reflecting the fact
that underlying dynamical mechanisms are entirely different, the
initial conditions determine not only the magnitude, but also the sign
of the net shift in baryonic number.

\section{Baryogenesis at preheating}

Dynamical principles of baryogenesis at preheating are relatively
simple (for detailed description of physical framework, Lagrangian and
dimensionless units used below see Ref.~\cite{bgks}). Energy transfer
from inflaton field into low-momentum modes of Higgs field results in
high rate of inequilibrium sphaleron transitions that appear very soon
after the beginning of resonance (see Figs. 8 and 14 of
paper~\cite{bgks}). On the other hand, in the course of resonance the
Higgs v.e.v.\ $\phii$ comes from zero to its post-resonant value close
to 1. Similar to earlier models of baryogenesis at first-order phase
transition (see e.g.~\cite{gts}), 
the time evolution of Higgs field gives rise to effective
chemical potential $\mue$ that appears through CP-breaking term $ -
\kappa |\phi|^2 \, \epsilon_{\mu\nu}F^{\mu\nu}$ in Lagrangian.  In the
presence of sphaleron transitions, this chemical potential drives the
Chern-Simons number $\ncs = {1 \over {2\pi}} \int A_1 dx$ in certain
direction (see Figs.~11 and 16 of Ref.~\cite{bgks} and
Fig.~\ref{bigres} below), thus producing the fermions. In our case,
the chemical potential is \cite{bgks,bg}
\be
\mue = - 4\pi\kappa\,\partial_0 \phii    \label{mueff}
\ee
and has nonzero value only while $\phii$ keeps changing.

The transition of $\phii$ to its stationary value isn't instant (see
Fig.~10 of ~\cite{bgks}); it takes about the same time as the resonance
itself, which is generally much slower than the first-order phase
transition. This is an important advantage of baryogenesis at
preheating over previous scenarios, because extended
baryoproduction time is essential for reducing possible destruction of
the created baryons by post-resonant sphaleron transitions, although
sphaleron freeze-out is related to Higgs thermalization and occurs on
different time scale.

\begin{figure}[t]
\centering
\hspace*{-6mm}
~~~\leavevmode\epsfysize=10cm \epsfbox{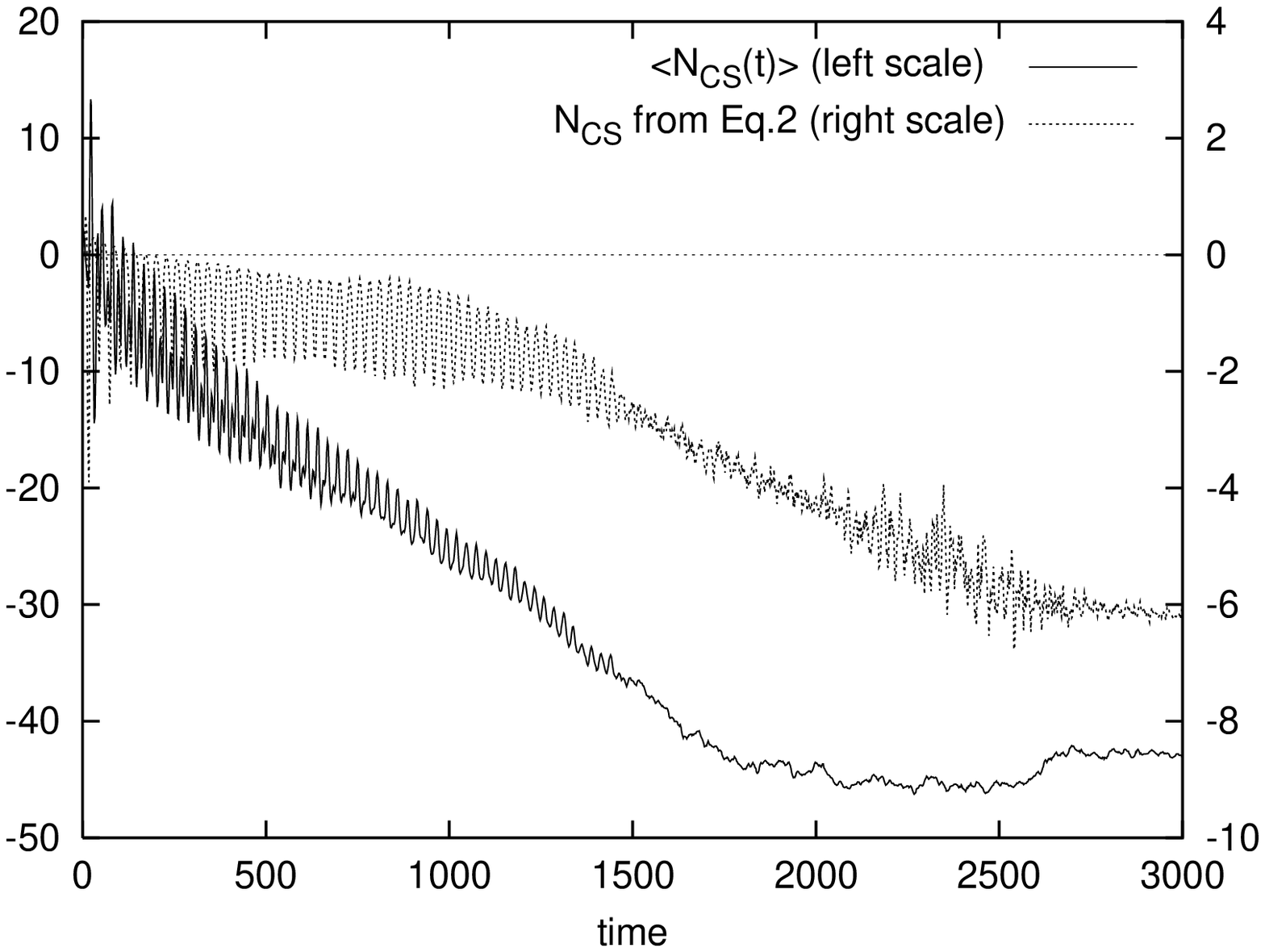} \\
\leavevmode\epsfysize=10cm \epsfbox{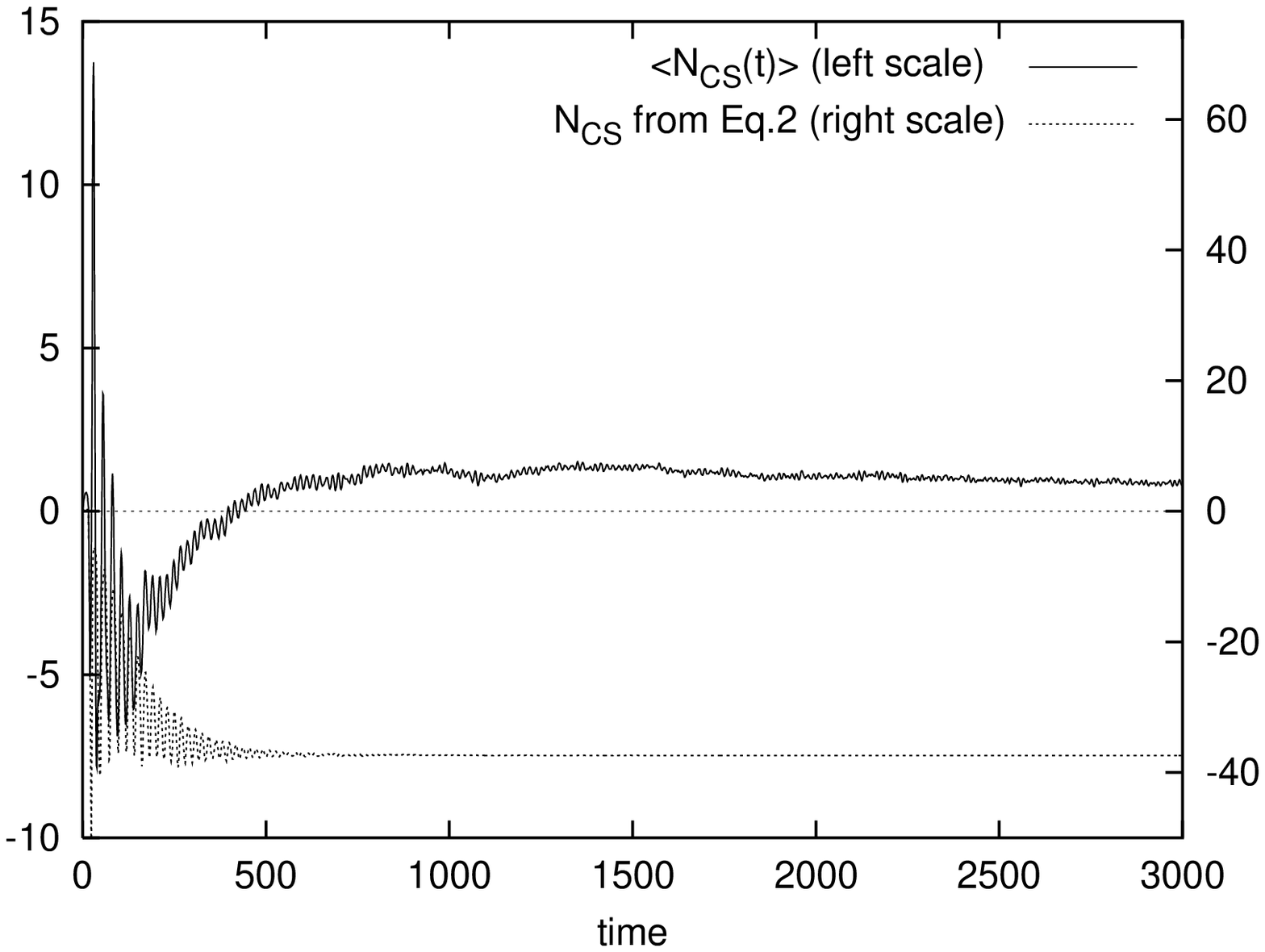}\\[3mm]
\caption{\label{bigres} (upper plot) Time evolution of Chern-Simons
number for $\kappa=-1$ and high  $\trh = 0.33\,v$ (solid line, identical to Fig.~11
of Ref.~\protect\cite{bgks}) compared to prediction of Eq.~(\ref{diffuse})
(dotted line).}
\end{figure}
\begin{figure}[!tbp]
\centering
\vspace*{-0.7cm}
\caption{\label{smallres} (lower plot) Same for reduced-energy runs
of~\protect\cite{bgks} with $\kappa=-1$ and $\trh = 0.094\,v$. On this plot $\ncst$, solid
line, differs from Fig.~16 of Ref.~\protect\cite{bgks} due to larger
statistics. Strong descripancy between observed $\ncst$ and that one predicted
by diffusion equation (\ref{diffuse}) means that the baryogenesis at
preheating, roughly described by Eq.~(\ref{diffuse}), is followed here
by some other baryoproduction mechanism --- baryogenesis after
preheating that moves $\ncs$ in opposite direction.}
\end{figure}

A common way of estimating the baryoproduction in this and similar
models is the use of diffusion equation for Chern-Simons number:
\be
{{d\ncs} \over {dt}} = -\gsph {\mue \over \teff} \label{diffuse}
\ee
where $\gsph$ is the rate of sphaleron transitions and
$\teff$ --- certain effective temperature. Note that Eq.~16
of~\cite{bgks} coincides with Eq.~(\ref{diffuse}) when $\Gamma_B = 0$
(the latter condition reflects the lack of fermionic backreaction in
our simulations). Once the $\phii$ value is increasing in the 
course of resonance, 
Eqs.~(\ref{mueff}) and (\ref{diffuse})
give the sign of $\Delta \ncs$:
\be
{{\Delta\ncs} \over \kappa} \propto -{\mue \over \kappa} \propto \partial_0
\phii  > 0 \label{sign} 
\ee

Eq.~(\ref{diffuse}) can be used for rough estimate of total generated
asymmetry by assuming final $\phii$ to be~1: 
\be 
\Delta \ncs \sim {4\pi\kappa \over \teff} \langle\gsph\rangle \label{delta}
\ee 
giving
$\Delta \ncs \sim -20$ for Fig.~\ref{bigres} and $\Delta \ncs \sim -30$ for
Fig.~\ref{smallres}. While the first estimate is correct within factor
of 2, the latter one has the wrong sign and is off by an order of
magnitude. This descripancy cannot be explained by simplifications
used for getting Eqs.~(\ref{sign}) and (\ref{delta}).

Indeed, it isn't obvious that the diffusion
equation Eq.~(\ref{diffuse}) can be reliably applicated to stronly
nonequilibrium dynamics of parametric resonance.  We check this
point directly by substituting $\langle\phi^*(t)\phi(t)\rangle$ and
$\Gamma(t)$ measured in numerical simulations~\cite{bgks} into
Eqs.~(\ref{mueff}) and (\ref{diffuse}),
taking $\teff=\trh$ and numerically integrating over time, see
Figs.~\ref{bigres} and \ref{smallres}. The only source of uncertainty in this
check is the effective temperature $\teff$ present in Eq.~(\ref{diffuse}). Exact
physical sense of this quantity isn't completely clear; however,
Figs.~6 and 13 of Ref.~\cite{bgks} show that various effective
temperatures stay rather close to final equilibrium value $\trh$ and
their time variations wouldn't considerably affect the time integral
of Eq.~(\ref{diffuse}).

As is clear from Fig.~\ref{bigres}, even for high reheating
temperatures the Eq.~(\ref{diffuse}) is adequate only for rough
qualitative analysis of $\ncs$ evolution. It cannot reproduce
correclty the short-time dynamics of $\ncs$ and provides no
explanation for the end of baryoproduction at time $\sim 1800$. Both
problems are unrelated to possible variations in $\teff$ which are
slow and disappear at time $\sim 1500$, see Fig.~6 of
Ref.~\cite{bgks}. For lower $\trh\sim 0.09\esph$ the diffusion
equation (\ref{diffuse}) is no longer able to give correct sign of
final asymmetry, see Fig.~\ref{smallres}. It is natural to suppose
that the evolution of $\ncs$ cannot be reduced to simple diffusion
described by Eq.~(\ref{diffuse}) and is sensitive to more subtle
dynamical effects. A good example of complicated dynamics beyond the
scope of Eq.~(\ref{diffuse}) is the baryogenesis after preheating.

\section{Baryogenesis after preheating}

The driving force for baryogenesis after preheating~\cite{bg} is the inflaton
field that keeps coherently oscillating after incomplete parametric
resonance (see Figs.~2 and 3 of Ref.~\cite{bg}). Inflaton zero mode
decays through thermalization which in our case is very slow. Its
oscillations considerably modify Higgs field effective potential
\be
V(\phi) = {\lambda\over 4}(|\phi|^2 - v^2)^2 + {1\over 2}g^2 \sigma^2
|\phi|^2
\ee
thus making inflaton zero mode $\ssqangle$ and Higgs v.e.v.\ $\phii$ 
to oscillate coherently. In its turn, $\phii$ oscillations generate
periodic CP-nonconservation through Eq.~(\ref{mueff}) and also 
trigger the sphaleron transitions that become synchronized to inflaton/Higgs 
oscillations. Interplay between these two
strongly correlated processes results into steady shift of $\ncs$ in
direction opposite to that one predicted by diffusion equation
(\ref{diffuse}). 

In this section we will analyze dynamics of Chern-Simons number $\ncs$
in the presence of oscillating $\mue$. Then, taking into account time
correlations to Higgs-triggered sphaleron transitions, we show why
$\ncs$ actually shifts in direction opposite to baryogenesis at
preheating.

\subsection{CP--$\phii$ correlations}

\begin{figure}[t]
\centering
\hspace*{-6mm}
\,~\leavevmode\epsfysize=10.21cm \epsfbox{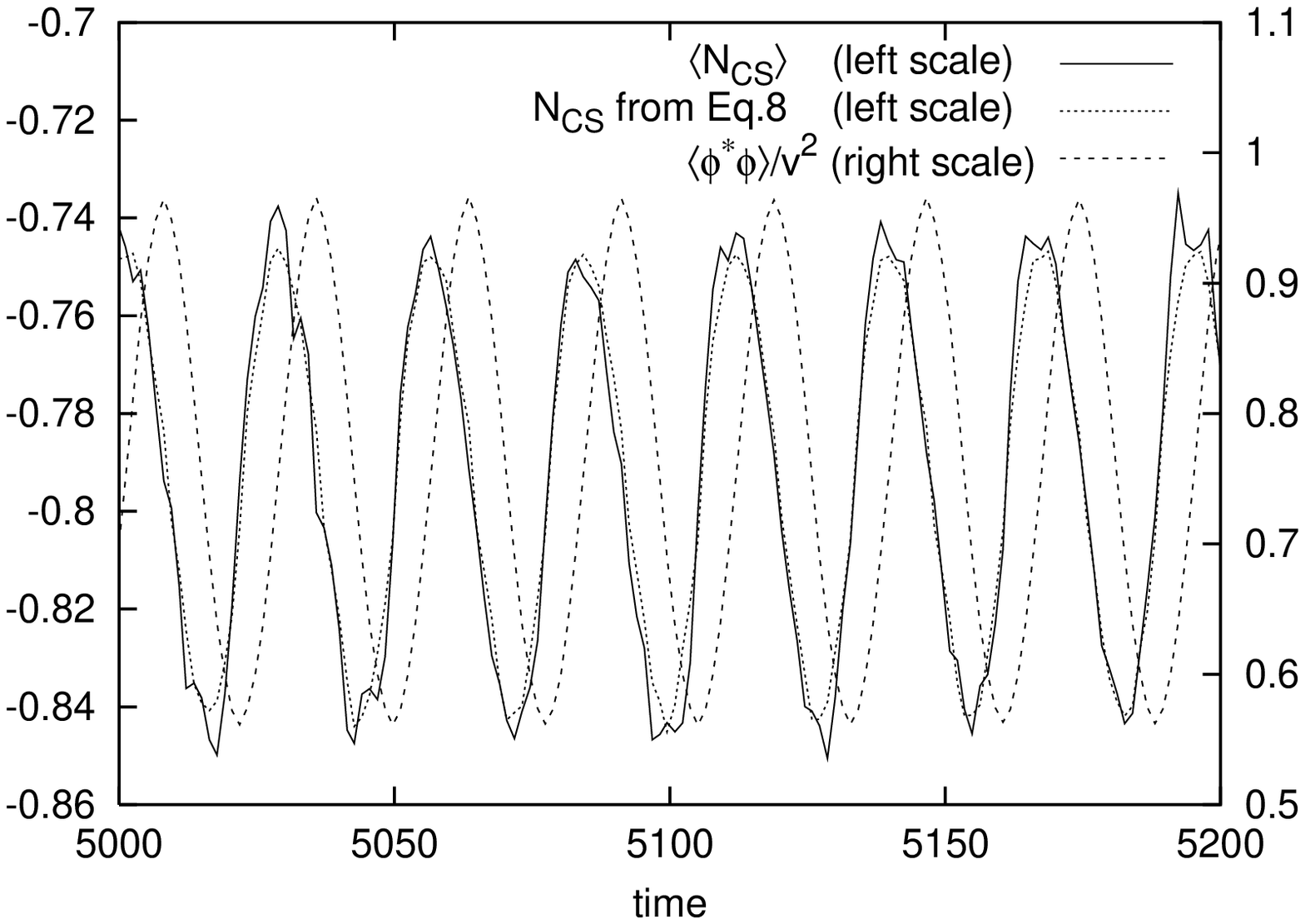} \\
\leavevmode\epsfysize=10cm \epsfbox{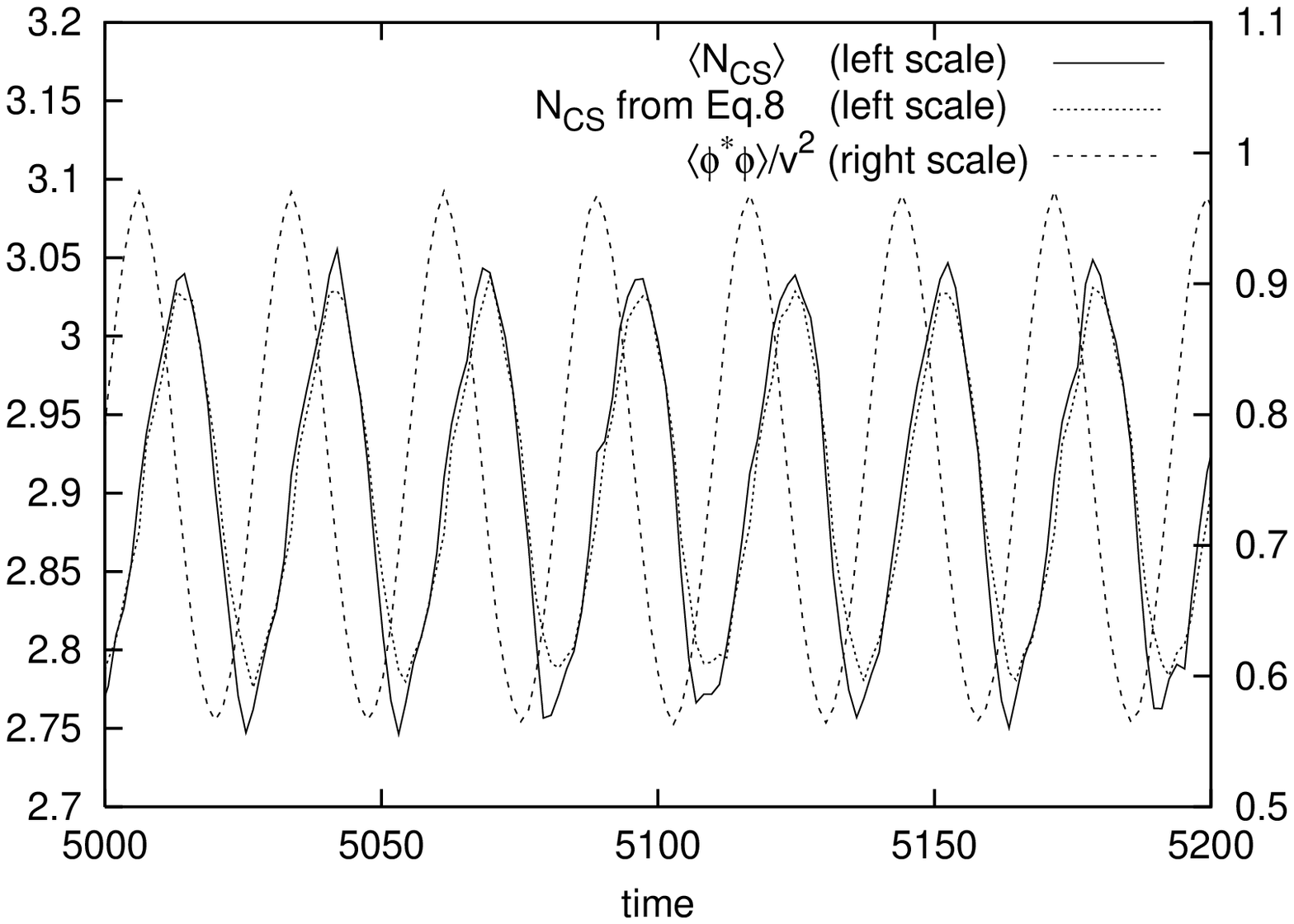}\\[3mm]
\caption{\label{kap_01} Comparison of phase relation between $\ncs$
and $\phii$ oscillations for $\kappa=0.1$ (upper plot) and
$\kappa=-0.25$ (lower plot) shows that their relative phase shift is
sensitive only to $\sign \kappa$. Both plots also confirm 
Eq.~(\ref{dncs}). The sphaleron transitions occur when $\phii$
is at minimum and $\ncs$ is close to its mean value $N_0$, see
text. Note that curves presented here are averaged over large ensemble
of independent runs, so $N_0$ has non-integer values $-0.8$ and
$2.9$, respectively.}
\end{figure}

Due to time derivative in $\mue$, Eq.~(\ref{mueff}),
periodical oscillations of $\phii$ clearly affect $\ncs$ behavior. For
$\kappa=0$, $\ncs$ is oscillating uncorrelated to $\phii$ with its own
characteristic frequency $\omega_0 = M_W \sim 1$ (these oscillations
are averaged out on the plots), while
introduction of non-zero $\kappa$ makes the $\ncs$ dynamics dominated
by $\mue$ even for relatively small $\kappa \sim 0.1$, see
Fig.~\ref{kap_01}. The amplitude of $\ncs$ oscillations at non-zero
$\kappa$ can be estimated using equation of motion~\cite{bgks,bg} for
$A_1$ field. Let's assume that $\ncs$ is oscillating around
its steady integer value $N_0$ which is the topological index of
current system vacuum. Making large gauge transformation to vacuum
with zero topological index and denoting the new Chern-Simons number
as $\dncs = \ncs - N_0$, one gets \be \ddot{\dncs} + 2\phii\dncs =
{L\over {2\pi}} \cdot 2\kappa\partial_0\phii
\label{dncs2}
\ee
(we assume the Higgs field to be homogeneous, i.e.~neglect all its
higher-momentum modes except the zero mode), and
\be 
\dncs = {{i\omega_\sigma} \over {{\omega_0}^2 -
{\omega_\sigma}^2}} \cdot {{\kappa L} \over \pi } \delta\phii
\label{dncs1} \ee 
Here $\delta\phii = Ae^{i\omega_\sigma t}$ (A is
the amplitude of $\phii$ oscillations), ${\omega_\sigma}^2 = 4 g^2\phii$
 is the frequency of inflaton oscillations, and
${\omega_0}^2 = 2\phii = {M_W}^2$ is the proper frequency of free
$\ncs$ oscillations not visible on Fig.~\ref{kap_01}. Once
$g^2\ll 1$, we can neglect ${\omega_\sigma}^2\ll {\omega_0}^2$ and
further simplify Eq.~(\ref{dncs2}):
\be
\dncs = {{\kappa L} \over {2\pi}} {{\partial_0\phii} \over {\phii}} =
- {L\over{4\pi^2}} {\mue \over \phii}
\label{dncs}
\ee

Both eqs.~(\ref{dncs1}) and (\ref{dncs}) result in the phase shift
between $\dncs$ and $\delta\phii$ being exactly ${\pi\over 2}
\sign\kappa$. This phase relation actually determines the sign of
baryonic asymmetry generated by $\Gamma-CP$ correlations.

\subsection{$\gsph$--$\phii$ correlations and baryoproduction}

Once $\phii$ is oscillating periodically, time averaging of $\mue$
(\ref{mueff}) and $\delta\ncs$ (\ref{dncs}) will give zero. In
our case this does not mean zero baryoproduction, because the
sphaleron transitions occur synchronously to the same oscillations of
$\phii$, thus favouring certain values of $\mue$.

The correlations between sphaleron transitions and $\phii$
oscillations are discussed in more detail in paper~\cite{bg}. It is
quite natural, however, that most transitions occur at $\phii$ close to its
minimum (see Figs.~4 and 6 of Ref.~\cite{bg}), once this corresponds
to minimal $\esph$. Note that at these moments $\mue \propto
\partial_0\phii =0$, so $\langle\mue\gsph\rangle=0$. In other words,
Eq.~(\ref{diffuse}) fails even after accounting for correlations between
$\mue$ and $\gsph$. Therefore, baryogenesis after preheating involves
dynamical effects that are time-nonlocal. Once most topological
transitions occur when $\phii$ is at minimum, $\phii$ value is decreasing 
during the whole
half-period of $\phii$ oscillation immediately preceeding the
transition, so  $\partial_0\phii < 0$ and $\mue$ has non-zero value of
certain sign. According to Eq.~(\ref{dncs}), $\dncs/\kappa
\propto \partial_0\phii < 0$ over the same half-period, see
Fig.~\ref{kap_01}. Therefore, the Chern-Simons number is always shifted
in certain direction for some period of time {\em just before} the
sphaleron transitions, making transitions with $\Delta \ncs =
-\sign\kappa$ more probable than transitions in opposite direction.

\begin{figure}[t]
\centering
\hspace*{-6mm}
~~ \leavevmode\epsfysize=10cm \epsfbox{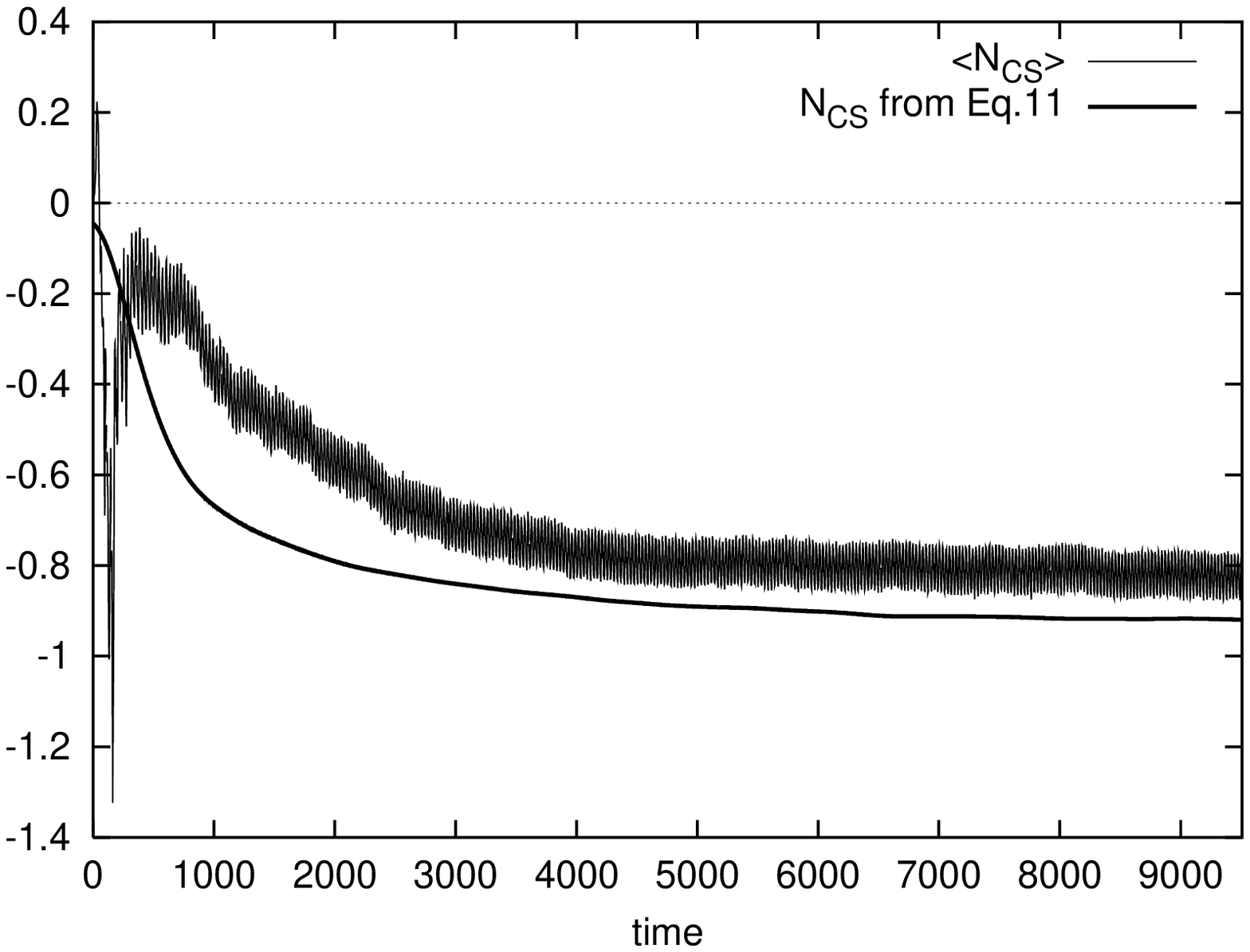} \\
\leavevmode\epsfysize=10cm \epsfbox{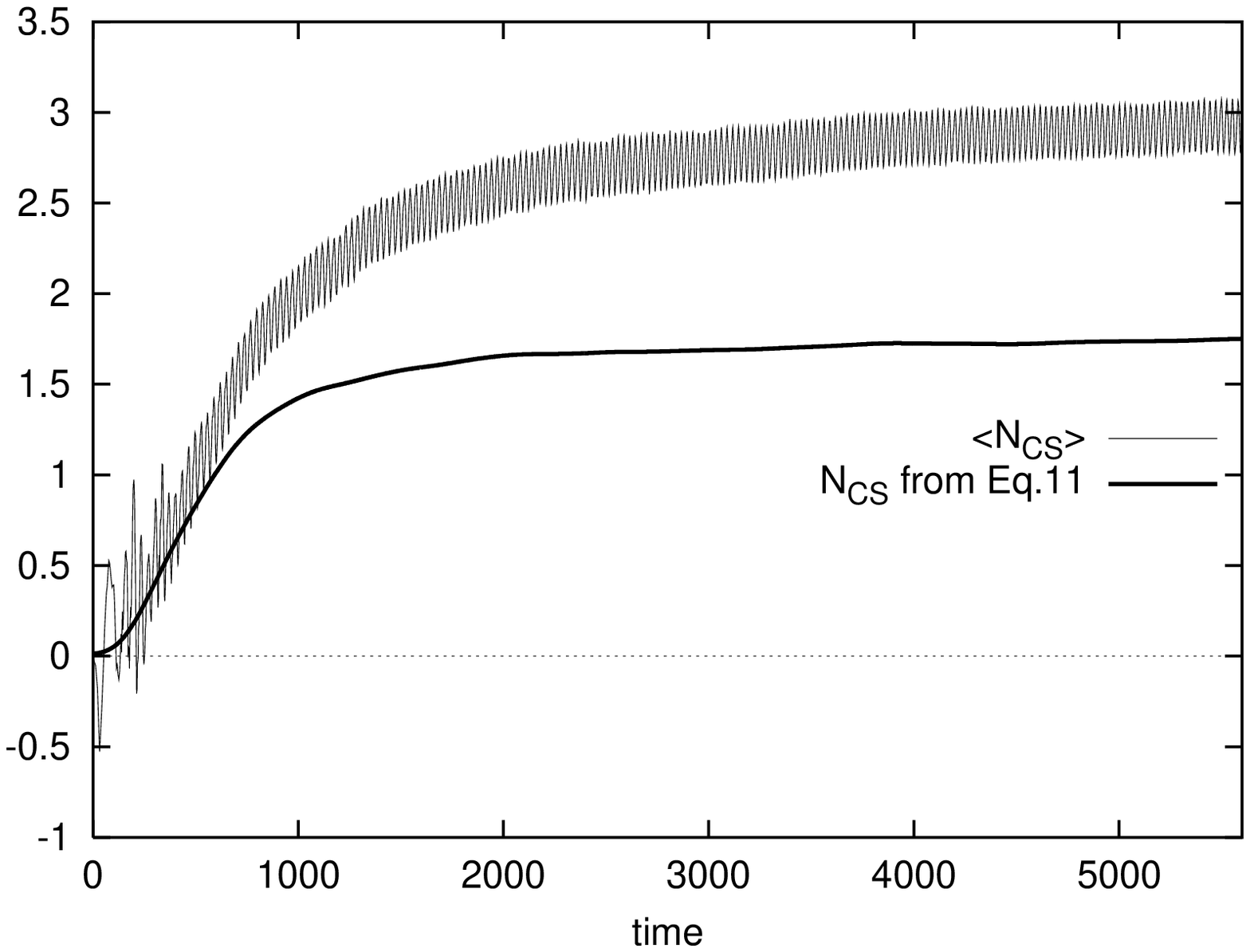}\\[3mm]
\caption{\label{compare} Eq.~(\ref{comp}) adequately describes dynamics
of Chern-Simons number both for $\kappa=0.1$ (upper plot) and
$\kappa=-0.25$ (lower plot), demonstrating the lack of wash-out for
baryogenesis after preheating. Unlike Fig.~\ref{smallres}, both
$\ncsa$ curves can be smoothly interpolated to origin, due to the
suppression of baryogenesis during preheating at very low
$\trh\sim10^{-2}v$.}
\end{figure}

This description  provides a simple estimate
for net asymmetry created by baryogenesis after preheating. Assuming
that every topological transitions effectively freezes the maximal
positive or negative
$\dncs$ that existed $1/4$-period before, one obtains from
Eq.~(\ref{dncs1}) (here we also use Eqs.~2.1 and 2.2 of Ref.~\cite{bg}):
\be
{{d\ncs} \over {dt}} = -\sign\kappa\cdot\left(\dncs\right)_{\rm max}\gsph =
- {{\kappa gvL}\over {2\pi}}  
\left({{\sigma_{\rm max}}\over{\sigma_c}}\right)^2\gsph \label{corr}
\ee
 and from Eq.~(\ref{dncs}):
\be
{{\Delta\ncs} \over \kappa} \sim \partial_0 \phii  < 0 \,. \label{sign2} 
\ee
(compare to Eq.~(\ref{sign})).
Substituting into Eq.~(\ref{corr}) parameter values from~\cite{bgks,bg},
one obtains 
\be
{{d\ncs} \over {dt}} \sim  -\,0.3\,\kappa\gsph \label{comp}
\ee
in reasonable agreement with numerical results, see Fig.~\ref{compare}.

Although baryogenesis after preheating is suppressed by inflaton-Higgs
coupling $g$, it keeps going as long as topological transitions occur,
see Eqs.~(\ref{corr}) and (\ref{comp}), and isn't affected by the wash-out
problem. It also shows how strongly the nontrivial dynamics can affect
simple estimates in the style of Eq.~(\ref{diffuse}).

\section{Conclusions}

Here we have shown that dynamical effects do play an important role in
baryogenesis at/after preheating, what is typical for complicated
nonequilibrium processes~\cite{ck,nauta}.  A simple (1+1)-dimensional
model of electroweak preheating-related baryogenesis, depending on
initial parameter values, may subsequently pass through two
dynamically distinct stages with different baryogenesis mechanisms
producing baryonic asymmetry of opposite signs. The second stage that
occurs after preheating is completely beyond the reach of
straightforward diffusion approach commonly used for analysis of
electroweak baryogenesis scenarios.

\section*{Acknowledgments}

The author is deeply indebted to his collaborators: Juan
Garc{\'\i}a-Bellido, Alexander Kusenko and Mikhail Shaposhnikov.
Presentation of these results at XXXV Rencontres de Moriond would be
impossible without kind support of conference organizers. This work is
supported in part by RBRF grant 98-02-17493a.

\end{document}